\documentclass[conference]{IEEEtran}
\IEEEoverridecommandlockouts

\usepackage{cite}
\usepackage{amsmath,amssymb,amsfonts}
\usepackage{algorithmic}
\usepackage{graphicx}
\usepackage{textcomp}
\usepackage{xcolor}
\usepackage{url}
\usepackage{booktabs}
\usepackage{newtxtext,newtxmath}

\def\BibTeX{{\rm B\kern-.05em{\sc i\kern-.025em b}\kern-.08em
    T\kern-.1667em\lower.7ex\hbox{E}\kern-.125emX}}

\begin{document}

\title{Evaluating Behavioral Context for Interpretable IAM Policy Risk Scoring in Cloud Environments}

\author{\IEEEauthorblockN{Yassin Elsharkawy}
\IEEEauthorblockA{\textit{Dept. of Computers and Artificial Intelligence} \\
\textit{Capital University}\\
Cairo, Egypt \\
ORCID: 0009-0007-2885-2546}
}

\maketitle

\bstctlcite{IEEEexample:BSTcontrol}

\begin{abstract}
IAM policy analysis typically emphasizes the authorization capabilities encoded in a policy, but security analyst review priority may also depend on the behavioral and environmental context surrounding a policy event. This paper evaluates whether contextual information provides measurable incremental value for interpretable IAM policy risk prioritization beyond policy and effective-authorization information. AWS is used as the experimental cloud provider because its IAM and audit-telemetry ecosystem enables controlled evaluation using AWS IAM Context Bench, a benchmark containing 534 real AWS experimental observations across policy, environment, and behavioral scenarios, including matched cases where policy and environment remain fixed while behavioral context changes. Three Explainable Boosting Machine models are evaluated under the same leakage-controlled grouped cross-validation protocol: a policy-centric baseline, a policy-plus-environment model, and a full-context model incorporating CloudTrail telemetry. The full-context model substantially reduces analyst-priority prediction error relative to the policy-centric baseline and closely tracks the reference priority ordering. In matched same-policy context pairs, the policy-centric model remains invariant, whereas the full-context model separates benign and suspicious behavioral conditions with high directional accuracy. The results also show improved concentration of high-priority cases at the top of simulated analyst review queues. These findings indicate that behavioral and environmental context can provide useful incremental information for analyst-oriented IAM risk prioritization while preserving an interpretable additive model structure. The formulation is applicable beyond AWS conceptually, although cross-provider validation remains future work.
\end{abstract}


\section{Introduction}
Cloud security operations routinely require security analysts to evaluate modifications to Identity and Access Management (IAM) configurations. In Amazon Web Services (AWS), IAM policies govern the authorization boundary of cloud tenants. Over-privileged policies, excessive permissions, and unintended escalation paths represent primary vectors for cloud security compromise \cite{bansal2024mitigating}. Consequently, existing literature and tooling predominantly emphasize static authorization analysis: parsing JSON permission documents, validating reachability, and auditing configurations against least-privilege principles \cite{backes2020stratified, dantoni2024automatically, eiers2022quantifying, parkinson2022identifying}.

While static analysis establishes what an identity is permitted to execute, authorization structure alone does not capture the operational circumstances under which an authorization event occurs. In enterprise security operations centers (SOCs), analysts triage dozens of policy modifications daily. An administrative role attached by a verified deployment pipeline presents vastly different operational urgency than the identical policy attached through an unmapped user-agent, without multi-factor authentication, preceded by reconnaissance activity \cite{akpara2025risk, chode2025ai}.

This distinction establishes two separate concepts: \emph{intrinsic policy risk}, representing the permissions and potential blast radius defined by authorization syntax; and \emph{analyst review priority}, representing the operational urgency with which an analyst should investigate the event given surrounding context. Behavioral context recorded in AWS CloudTrail audit logs does not alter the underlying authorization granted by AWS. Rather, it provides circumstantial evidence regarding how authorization is exercised \cite{aws_cloudtrail_unexpected, adediran2026retrieval}.

\begin{figure}[htbp]
\centering
\includegraphics[width=0.86\columnwidth]{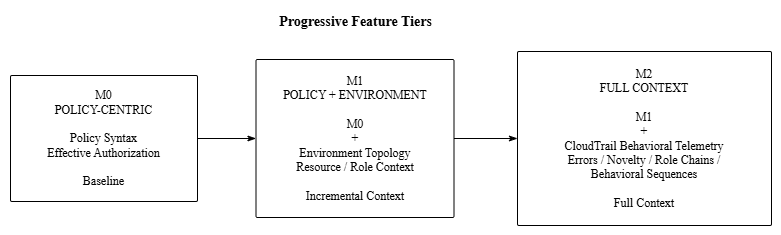}
\caption{End-to-end evaluation workflow: IAM policy events and infrastructure states are captured with CloudTrail telemetry, split via grouped cross-validation, and scored by an additive EBM to produce analyst-priority rankings.}
\label{fig:workflow}
\end{figure}

Although contextual access control and log-based threat detection have been studied separately \cite{atlam2020risk, pamungkas2026contextual, abibulaiev2026context}, an open empirical question remains: Does AWS CloudTrail behavioral context provide measurable incremental value for analyst-oriented IAM policy risk prioritization beyond policy-centric and environmental information? More specifically, when policy syntax and effective authorization are held constant, can an interpretable model reliably distinguish benign operational events from suspicious activity?

To answer this question, this paper evaluates \textsc{AWS-IAM-ContextBench}, an empirical benchmark comprising 534 validated real AWS observations across 13 scenario families. The benchmark features 208 matched Type-A pairs wherein the policy document, environment topology, and effective authorizations remain fixed while CloudTrail behavioral context varies.

We evaluate three progressively enriched tiers of an Explainable Boosting Machine (EBM) under a leakage-controlled grouped cross-validation protocol: the policy-centric baseline $M_0$ (15 features), the environment-augmented model $M_1$ (28 features), and the full-context model $M_2$ (42 features). Fig.~\ref{fig:workflow} outlines the evaluation pipeline.

The primary contributions of this study are:
A. \emph{Incremental Ablation Evaluation}: We empirically measure the value of environment and CloudTrail behavioral telemetry, observing an MAE reduction of 74.0\% ($p < 10^{-15}$) and a Spearman correlation increase to $\rho = 0.9726$.
B. \emph{Controlled Matched-Pair Sensitivity Analysis}: Using 207 verified Type-A pairs, we demonstrate that while policy-centric scoring remains invariant ($\Delta = 0.0000$), context-aware scoring distinguishes suspicious from benign activity with 96.62\% direction accuracy ($p = 3.68 \times 10^{-30}$).
C. \emph{Interpretable Triage Assessment}: We evaluate simulated review queues, observing up to a +19.7\% lift in NDCG@10 while preserving additive feature attribution.

\section{Related Work}

\subsection{IAM Policy Analysis and Authorization}
Static IAM policy analysis focuses on formal specification and least-privilege verification. Backes et al. introduced stratified abstraction and automated reasoning for AWS policies \cite{backes2020stratified}. D'Antoni et al. investigated automated privilege reduction for cloud access policies \cite{dantoni2024automatically}, while Eiers et al. developed techniques to quantify policy permissiveness \cite{eiers2022quantifying}. Hu et al. examined hybrid privilege escalation detection in cloud IAM \cite{hu2023tac}, Bansal analyzed risk mitigation via policy controls \cite{bansal2024mitigating}, and Parkinson and Khana applied fuzzy logic to detect over-entitlement \cite{parkinson2022identifying}. These methods evaluate policy documents in isolation, remaining uninformative regarding runtime behavioral circumstances surrounding policy deployment.

\subsection{Context-Aware and Risk-Adaptive Access Control}
Risk-Adaptive Access Control (RAdAC) dynamically adjusts authorization decisions based on operational risk, environmental state, and situational factors \cite{atlam2020risk, alharbe2023access}. Akpara examined risk-weighted access control combining least privilege with authentication signals \cite{akpara2025risk}. Chode analyzed AI-driven risk-adaptive authorization for multi-tenant cloud APIs \cite{chode2025ai}. Kulkarni et al. proposed dynamic trust-weighted attribute-based access control \cite{kulkarni2025dtw}, Feng et al. explored context-aware decisions using language models \cite{feng2026static}, and Piriaei et al. formulated deep sequential risk-adaptive architectures \cite{piriaei2025real}. While RAdAC addresses dynamic runtime enforcement, enterprise cloud platforms enforce binary IAM evaluation. Our study evaluates post-event triage: prioritizing policy events for human security review without modifying runtime enforcement.

\subsection{Cloud Audit Telemetry and Behavioral Analytics}
AWS CloudTrail provides an immutable record of account activity \cite{aws_cloudtrail_unexpected}. Adediran et al. mapped CloudTrail events to MITRE ATT\&CK tactics \cite{adediran2026retrieval}, Pamungkas and Sutopo evaluated unsupervised contextual anomaly detection on CloudTrail logs \cite{pamungkas2026contextual}, and Abibulaiev et al. developed real-time anomaly detection pipelines for cloud API traffic \cite{abibulaiev2026context}. Madireddy explored graph neural networks for cloud IAM logs \cite{madireddy2025graph}, Alkharabsheh et al. developed insider-threat frameworks with explainable containment \cite{alkharabsheh2026artificial}, Konakalla et al. surveyed AI-driven cloud monitoring \cite{konakalla2026ai}, and Markande et al. examined SaaS API monitoring for access anomalies \cite{markande2026saas}.

Existing work frequently focuses on authorization policies, identities, access requests, or behavioral activity as separate analysis objects. This study instead evaluates the IAM policy event as an analyst-facing prioritization object, isolating the incremental contribution of CloudTrail behavioral context using matched same-policy observations.

\section{Problem Formulation}
Let an IAM policy observation be represented by a feature vector partitioned into three distinct tiers:
\begin{equation}
x = [x_p, x_e, x_c] \in \mathcal{X}
\label{eq:features}
\end{equation}
where $x_p \in \mathbb{R}^{d_p}$ denotes policy syntax and effective authorization metrics, $x_e \in \mathbb{R}^{d_e}$ represents infrastructure environment topology, and $x_c \in \mathbb{R}^{d_c}$ encapsulates CloudTrail behavioral telemetry.

\begin{figure}[htbp]
\centering
\includegraphics[width=0.86\columnwidth]{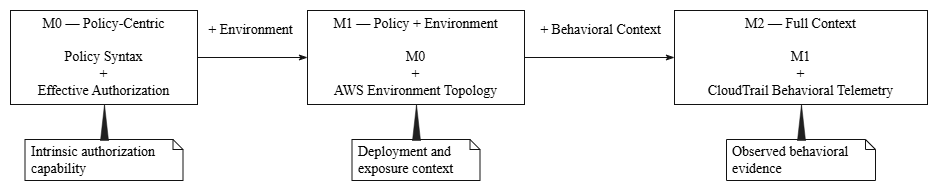}
\caption{Progressive feature tiers: $M_0$ (policy authorization), $M_1$ (environment topology), and $M_2$ (CloudTrail behavioral telemetry).}
\label{fig:ablation}
\end{figure}

Let $y \in \mathbb{R}^+$ denote the reference analyst-priority score assigned by an independent oracle. We define three progressive scoring models:
\begin{align}
M_0(x_p) &= f_0(x_p), \label{eq:m0} \\
M_1(x_p, x_e) &= f_1(x_p, x_e), \label{eq:m1} \\
M_2(x_p, x_e, x_c) &= f_2(x_p, x_e, x_c). \label{eq:m2}
\end{align}
The progressive feature tiering is illustrated in Fig.~\ref{fig:ablation}.

The objective is to estimate analyst review priority rather than predicting authorization outcomes. To evaluate behavioral sensitivity independently of policy confounding, we formalize the matched-pair condition. Let $(i, j)$ denote a Type-A pair of observations such that:
\begin{equation}
x_p^{(i)} = x_p^{(j)}, \qquad x_e^{(i)} = x_e^{(j)}, \qquad x_c^{(i)} \neq x_c^{(j)}
\label{eq:paired}
\end{equation}
where observation $i$ corresponds to an event executed under suspicious behavioral context and observation $j$ represents the identical policy under normal baseline activity.

Because the policy and environment are identical, a policy-centric model is constrained to output an invariant score:
\begin{equation}
M_0(x_p^{(i)}) - M_0(x_p^{(j)}) \approx 0
\label{eq:m0_inv}
\end{equation}
Conversely, a context-aware model $M_2$ is expected to produce a positive priority gap:
\begin{equation}
\Delta_{ij} = M_2(x^{(i)}) - M_2(x^{(j)}) > 0
\label{eq:gap}
\end{equation}
The magnitude and direction of $\Delta_{ij}$ quantify the sensitivity of the model to behavioral evidence.

\begin{table}[htbp]
\caption{AWS-IAM-ContextBench Benchmark Composition}
\label{tab:dataset}
\centering
\footnotesize
\setlength{\tabcolsep}{3pt}
\begin{tabular}{@{}lcp{4.0cm}@{}}
\hline
Dimension & Count & Description / Experimental Property \\
\hline
Total Samples & 534 & Controlled real-AWS experimental runs \\
Type-A Pairs & 208 & 416 samples; matched policy and env. \\
Type-B Pairs & 39 & 78 samples; matched policy, differing env. \\
Unpaired Samples & 40 & Edge-case structural policy variants \\
Scenario Families & 13 & S01--S13 privilege escalation / exfil. \\
Policy Lineages & 96 & Grouping units for 5-fold CV \\
AWS Environments & 16 & Distinct infrastructure topologies \\
Behavior Patterns & 12 & Real CloudTrail sequences (5 benign, 7 attack) \\
Primary Target & -- & Reference priority score ($\mu=20.29, \sigma=7.84$) \\
\hline
\end{tabular}
\end{table}

\section{AWS-IAM-ContextBench Dataset}

\subsection{Benchmark Composition and Invariants}
The empirical evaluation is conducted on \textsc{AWS-IAM-ContextBench} (v3.0), generated from controlled executions in a dedicated AWS research account. The benchmark comprises $N=534$ validated experimental observations. Table~\ref{tab:dataset} outlines the dataset composition.

The benchmark contains 208 Type-A matched pairs (416 samples) constructed to isolate behavioral context. Each pair shares identical canonical policy documents, identical effective authorizations, and identical infrastructure topologies, differing primarily in actor authentication, error activity, and API call sequences. Furthermore, 39 Type-B pairs (78 samples) hold policy and behavioral context fixed while varying the environment, and 40 unpaired samples provide coverage of additional structural variants.

The benchmark covers 13 scenario families across three categories: privilege escalation (S01 AttachRolePolicy, S02 CreatePolicyVersion, S03 SetDefaultPolicyVersion, S04 PassRole+Compute, S12 Cross-Principal Escalation); data exfiltration (S05 S3 Bucket Policy, S06 KMS Key Policy, S07 Secrets Manager, S08 SSM Parameter Exposure); and governance and boundary changes (S09 Policy Version Rollback, S10 Trust Policy Expansion, S11 Permissions Boundary Removal, S13 Least-Privilege Remediation).

The selection of these scenario families was informed by previously documented IAM privilege-escalation and access-control security scenarios, including TAC: Hybrid IAM Privilege Escalation Detection \cite{hu2023tac} and CloudFix, which studies real cloud access-control policy repair cases \cite{Hall2025CloudFix}. Additional publicly documented security research, laboratory scenarios, and cloud threat cases were also reviewed to identify representative privilege-escalation, data-exposure, policy-governance, and defense-evasion conditions. These sources were used in scenario design rather than to reproduce their datasets or experimental procedures directly.

The resulting benchmark is designed to approximate enterprise-style IAM activity and infrastructure configurations while retaining the control required for matched experimental evaluation. In particular, the environment includes multiple AWS services, policy histories, identity and role relationships, and behavioral sequences representative of security-relevant cloud operations. This controlled construction enables the study to isolate contextual effects that would be difficult to separate in an unconstrained production environment.

As a supplementary analysis of the frozen out-of-fold predictions, 69 of the 79 structural policy hashes occur in only one cross-validation fold, yielding 371 observations whose structural policy hash is absent from the corresponding training fold. On this subset, $M_2$ retains a Spearman correlation of $\rho=0.9674$ and MAE of 1.2236, compared with $\rho=0.2506$ and MAE of 4.8427 for $M_0$. This provides additional evidence that the contextual model retains strong prioritization performance when the observed policy structure is not represented in the corresponding training data.

\subsection{Reference Target Variable}
The dependent variable is \texttt{gt\_analyst\_priority\_score}, a continuous oracle score ($\mu = 20.29, \sigma = 7.84$, range $[4.60, 53.90]$) reflecting investigative urgency. The score is computed by an independent deterministic oracle evaluating authorization breadth, blast radius, authentication novelty, and activity timing. The target is never exposed to input features. A secondary metric, \texttt{gt\_security\_risk\_score}, measures intrinsic permission severity; across all verified Type-A pairs, intrinsic security risk is identical between benign and suspicious peers ($\Delta = 0.0$), confirming that reference priority variation arises from behavioral evidence.

\begin{table}[htbp]
\caption{Model Definitions and Feature Tiers}
\label{tab:models}
\centering
\footnotesize
\setlength{\tabcolsep}{3pt}
\begin{tabular}{@{}llcp{4.0cm}@{}}
\hline
Model & Tier & Feat. & Summary of Included Feature Signals \\
\hline
$M_0$ & Policy & 15 & Syntax, wildcards, statements, effective auth \\
$M_1$ & Env. & 28 & $M_0$ + Admin roles, KMS/Secrets, boundaries \\
$M_2$ & Context & 42 & $M_1$ + CloudTrail errors, user-agent, MFA \\
\hline
\end{tabular}
\end{table}

\subsection{Feature Quality and Leakage Prevention}
Feature engineering enforces strict leakage boundaries. First, 52 provenance, identifier, and metadata fields are blacklisted, including sample and pair IDs, policy hashes, AWS region, and raw timestamps. Second, all ground-truth fields are excluded from inputs. Third, variance auditing deterministically removed four zero-variance features (\texttt{policy\_default\_version}, \texttt{time\_deviation}, \texttt{first\_time\_action}, \texttt{actor\_deviation}), leaving 42 active features for $M_2$. Finally, raw JSON policy text is excluded in favor of semantic authorization metrics.

\subsection{Leakage-Controlled Grouped Cross-Validation}
To prevent data leakage across paired samples and shared policy lineages, all models are evaluated via 5-fold \texttt{StratifiedGroupKFold} grouped strictly on \texttt{policy\_lineage\_id}. The 534 observations originate from 96 unique policy lineages. Stratification is maintained across scenario families.

This protocol enforces three invariants: no policy lineage crosses between training and validation folds; both members of every Type-A and Type-B pair are co-located in the same fold; and preprocessing statistics are computed strictly on training partitions. All reported metrics derive from out-of-fold (OOF) predictions.

\section{Interpretable Risk Scoring Method}

\subsection{Explainable Boosting Machines}
To ensure that prioritization decisions remain transparent to security analysts, we implement Explainable Boosting Machines (EBM) \cite{nori2019interpretml}. EBM is an implementation of Generalized Additive Models (GAMs) trained via cyclic gradient boosting:
\begin{equation}
\hat{y} = \beta_0 + \sum_{k=1}^{D} f_k(x_k)
\label{eq:ebm}
\end{equation}
where $\beta_0$ is the intercept and each $f_k(x_k)$ is a non-parametric univariate shape function learned through bagged shallow regression trees.

We enforce $\text{interactions} = 0$, restricting the model to strictly additive main effects. This design ensures modular interpretability: the predicted review priority equals the sum of independent feature contributions, enabling analysts to inspect the exact numerical contribution of policy, environment, and CloudTrail features.

\subsection{Model Configuration}
Table~\ref{tab:models} summarizes the model definitions. Models $M_0$, $M_1$, and $M_2$ use identical hyperparameters: $\text{max\_bins} = 32$, $\text{outer\_bags} = 8$, $\text{learning\_rate} = 0.03$, $\text{max\_rounds} = 2000$, and random seed 42, evaluated across identical folds.

\begin{table}[htbp]
\caption{Experiment 1: Incremental Context Ablation Results (Out-of-Fold Cross-Validation, $N=534$)}
\label{tab:ablation}
\centering
\footnotesize
\setlength{\tabcolsep}{3pt}
\begin{tabular}{@{}lcccc@{}}
\hline
Metric & $M_0$ (Pol) & $M_1$ (Env) & $M_2$ (Ctx) & $\Delta(M_2 - M_0)$ \\
\hline
MAE $\downarrow$ & 4.8859 & 4.5272 & 1.2702 & -3.6157 \\
RMSE $\downarrow$ & 6.0335 & 5.2609 & 1.7286 & -4.3048 \\
Spearman $\rho$ $\uparrow$ & 0.4034 & 0.5666 & 0.9726 & +0.5692 \\
Kendall $\tau$ $\uparrow$ & 0.2835 & 0.4072 & 0.8640 & +0.5805 \\
NDCG@10 $\uparrow$ & 0.8446 & 0.8714 & 0.9787 & +0.1342 \\
NDCG@25 $\uparrow$ & 0.9166 & 0.9321 & 0.9874 & +0.0708 \\
NDCG@50 $\uparrow$ & 0.8686 & 0.9343 & 0.9864 & +0.1178 \\
\hline
\end{tabular}
\end{table}

\section{Experiments and Ablations}

\subsection{Experiment 1: Incremental Context Ablation}
Experiment 1 evaluates whether incorporating environment topology ($M_1$) and CloudTrail telemetry ($M_2$) yields measurable improvements over the policy-centric baseline ($M_0$) across all 534 out-of-fold samples.

\begin{figure}[htbp]
\centering
\includegraphics[width=0.86\columnwidth]{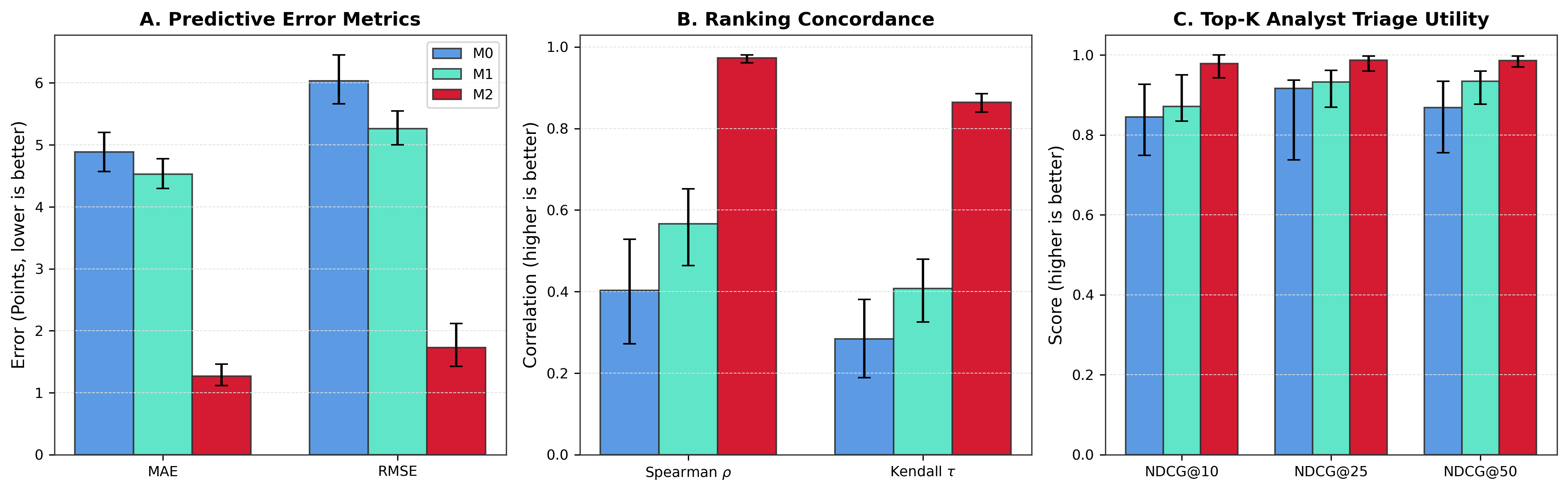}
\caption{Out-of-fold performance comparison of $M_0$, $M_1$, and $M_2$. Error bars represent 95\% cluster-bootstrap confidence intervals.}
\label{fig:main_results}
\end{figure}

Table~\ref{tab:ablation} and Fig.~\ref{fig:main_results} report out-of-fold performance metrics. The policy-centric baseline $M_0$ achieves an MAE of 4.8859 and an RMSE of 6.0335, with Spearman correlation $\rho = 0.4034$. Adding environment features in $M_1$ yields modest gains, reducing MAE to 4.5272 ($\Delta = -0.3587$) and increasing $\rho$ to 0.5666 ($\Delta = +0.1632$).

Incorporating CloudTrail behavioral telemetry in $M_2$ produces a substantial performance improvement: MAE decreases by 74.0\% to 1.2702 ($\Delta = -3.6157$, 95\% CI: $[-3.92, -3.29]$); RMSE decreases by 71.4\% to 1.7286 ($\Delta = -4.3048$, 95\% CI: $[-4.63, -3.97]$); Spearman's $\rho$ rises from 0.4034 to 0.9726 ($\Delta = +0.5692$, 95\% CI: $[+0.45, +0.70]$), indicating close tracking of reference priority; and Kendall's $\tau$ increases from 0.2835 to 0.8640 ($\Delta = +0.5805$). A paired Wilcoxon signed-rank test on absolute errors between $M_2$ and $M_0$ confirms that the reduction in error is statistically significant ($W = 0.0, p < 10^{-15}$).

\begin{figure}[htbp]
\centering
\includegraphics[width=0.86\columnwidth]{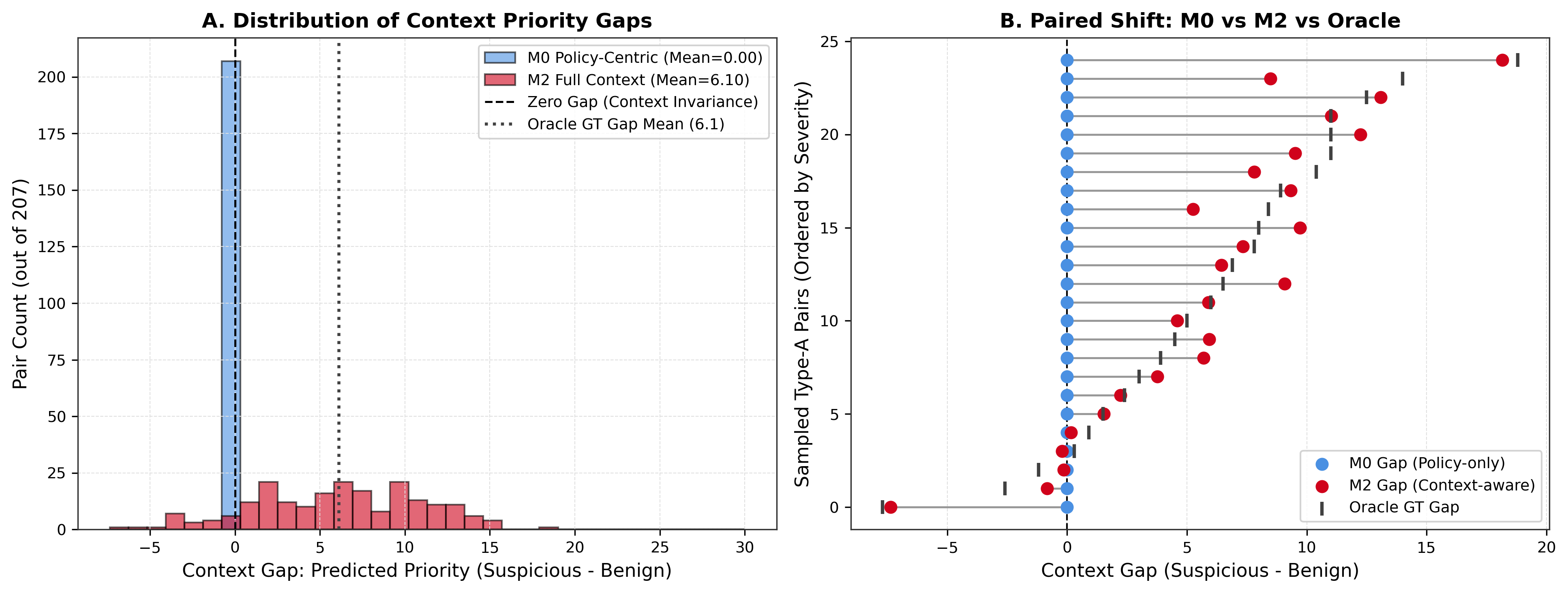}
\caption{Matched Type-A context pair separation ($N=207$). $M_0$ predictions remain invariant ($\Delta = 0.0000$), whereas $M_2$ separates suspicious contexts from benign baselines (+6.1049 mean gap).}
\label{fig:context_pairs}
\end{figure}

\begin{table}[htbp]
\caption{Experiment 2: Matched Type-A Context Pair Evaluation ($N=207$ Pairs)}
\label{tab:pairs}
\centering
\footnotesize
\setlength{\tabcolsep}{4pt}
\begin{tabular}{@{}p{4.0cm}cc@{}}
\hline
Evaluation Metric & $M_0$ (Policy) & $M_2$ (Full Ctx) \\
\hline
Mean Context Priority Gap ($\Delta_{ij}$) & 0.0000 & +6.1049 \\
Median Context Priority Gap & 0.0000 & +6.3517 \\
Mean Absolute Invariance Gap & 0.0000 & 6.1049 \\
Pairwise Direction Accuracy & 0.0\% & 96.62\% \\
Pairwise Ranking Accuracy & 0.0\% & 89.37\% \\
Context Gap MAE (vs. Oracle) & 6.6657 & 1.1051 \\
Gap Spearman Correlation & 0.0000 & 0.9573 \\
\hline
\end{tabular}
\end{table}

\subsection{Experiment 2: Same-Policy / Different-Context Paired Test}
Experiment 2 investigates whether $M_2$ detects behavioral shifts when authorization is held constant. Prior to evaluation, pair validation was conducted across all 208 Type-A pairs. One pair (\texttt{p\_s01\_004}) exhibited a single-action discrepancy in \texttt{effective\_denied\_action\_count} and was quarantined, leaving 207 strictly verified pairs.

Table~\ref{tab:pairs} and Fig.~\ref{fig:context_pairs} summarize the paired results. For $M_0$, the mean and median predicted context gaps are 0.0000 ($\text{MAE} = 6.6657$ against oracle gap). Because policy syntax and effective authorizations are identical, $M_0$ cannot differentiate the events.

In contrast, $M_2$ achieves a mean context gap of +6.1049 points (95\% CI: $[+5.43, +6.74]$), aligning closely with the oracle reference gap ($\mu = 6.1092$); a pairwise direction accuracy of 96.62\% (200 of 207 pairs correctly prioritized the suspicious event above its benign peer; 95\% CI: $[93.72\%, 98.55\%]$); and a gap rank correlation of $\rho = 0.9573$ ($p < 10^{-15}$). A paired Wilcoxon signed-rank test confirms that the context gap produced by $M_2$ is statistically significant ($W = 105.0, p = 3.68 \times 10^{-30}$). A paired permutation test with 5,000 resamples yields $p < 0.0002$, with a standardized effect size of $\text{Cohen's } d = 1.2389$.

\begin{table}[htbp]
\caption{Experiment 3: Analyst Review Queue Triage Performance Across Inspection Budgets ($K$)}
\label{tab:triage}
\centering
\footnotesize
\setlength{\tabcolsep}{3pt}
\begin{tabular}{@{}clcccc@{}}
\hline
$K$ & Metric & $M_0$ & $M_1$ & $M_2$ & Lift vs. $M_0$ (\%) \\
\hline
10 & NDCG@10 & 0.8176 & 0.8743 & 0.9787 & +19.7\% \\
10 & Precision@10 & 0.9000 & 0.7000 & 1.0000 & +11.1\% \\
10 & Recall@10 & 0.3000 & 0.2333 & 0.3333 & +11.1\% \\
\hline
25 & NDCG@25 & 0.9027 & 0.9336 & 0.9874 & +9.4\% \\
25 & Precision@25 & 0.8000 & 0.8000 & 0.9600 & +20.0\% \\
25 & Recall@25 & 0.6667 & 0.6667 & 0.8000 & +20.0\% \\
\hline
50 & NDCG@50 & 0.8616 & 0.9353 & 0.9864 & +14.5\% \\
50 & Precision@50 & 0.4600 & 0.6000 & 0.5600 & +21.7\% \\
50 & Recall@50 & 0.7667 & 1.0000 & 0.9333 & +21.7\% \\
\hline
\end{tabular}
\end{table}

\subsection{Experiment 3: Analyst Triage Utility}
Experiment 3 examines the practical utility of context scoring under constrained review budgets. Out-of-fold predictions from $M_0$, $M_1$, and $M_2$ are sorted to simulate a prioritized review queue. Table~\ref{tab:triage} reports Normalized Discounted Cumulative Gain (NDCG@K), Precision@K, and Recall@K for $K \in \{10, 25, 50\}$.

For a triage budget of $K=10$ alerts, $M_2$ achieves an NDCG@10 of 0.9787 compared to 0.8176 for $M_0$, representing a +19.7\% triage lift. Furthermore, $M_2$ achieves Precision@10 = 1.0000 on the evaluated benchmark. At $K=25$, $M_2$ achieves 0.9600 Precision@25 and 0.8000 Recall@25, outperforming $M_0$ (0.8000 Precision, 0.6667 Recall) by +20.0\% in both metrics. These results indicate that context-aware scoring concentrates high-priority cases toward the front of the review queue, which may improve triage efficiency in constrained security operations.

\subsection{Multi-Seed Robustness}
To confirm that results do not depend on random initialization, the full pipeline was re-evaluated across seeds 2026 and 31415. Results remained consistent: $\Delta(M_2 - M_0)$ MAE was $-3.6157$ (seed 42), $-3.6348$ (seed 2026), and $-3.6454$ (seed 31415); Type-A direction accuracy was 96.62\% (seed 42), 97.58\% (seed 2026), and 97.10\% (seed 31415); mean context gap was $+6.1049$ (seed 42), $+6.1506$ (seed 2026), and $+6.1300$ (seed 31415).

\section{Analyst Explanations and Case Studies}

\subsection{Additive Feature Attribution}
\begin{figure}[htbp]
\centering
\includegraphics[width=0.86\columnwidth]{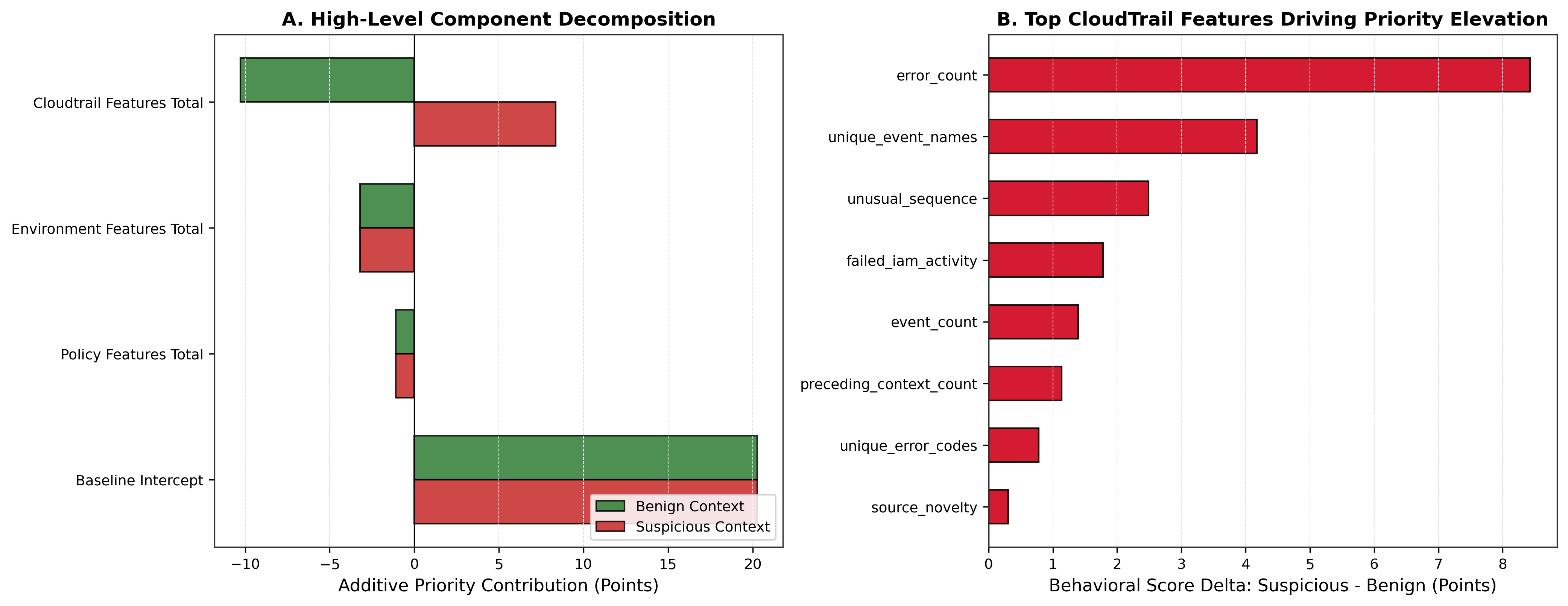}
\caption{Local additive EBM explanation for matched Type-A pair \texttt{p\_s01\_002}. Anomalous CloudTrail telemetry elevates priority by $+7.79$ points while policy and environment contributions remain identical.}
\label{fig:explanations}
\end{figure}

Because EBM enforces $\text{interactions}=0$, an analyst can decompose any prediction into explicit feature scores \cite{nori2019interpretml, fernandez2020explaining, nandi2026hierarchical}. Fig.~\ref{fig:explanations} illustrates this decomposition for a matched Type-A pair from scenario family S01 (\texttt{AttachRolePolicy} privilege escalation).

In both instances, the IAM policy grants administrative permissions (\texttt{AdministratorAccess}) within identical infrastructure environments. The model assigns identical baseline terms: intercept $\beta_0 = 20.30$, policy features $\sum f_p = -2.78$, and environment features $\sum f_e = -0.49$. Under benign execution (left), an established role via the AWS console with MFA yields $\sum f_c = -0.71$ (final priority 16.32). Under suspicious execution (right), the identical policy attached via an unmapped user-agent without MFA following failed IAM enumeration attempts contributes $\sum f_c = +7.08$, elevating priority to 24.11 ($\Delta = +7.79$). Because policy and environment terms are matched, the prediction delta arises entirely from CloudTrail telemetry.

\subsection{Diagnostic Error Analysis}
Auditing representative errors reveals that $M_2$ under-predictions primarily occur under telemetry attenuation: when an actor deliberately paces API calls or operates within services where CloudTrail delivery is delayed, causing $x_c$ to resemble a baseline state. Conversely, false positives occurred when benign administrators performed legitimate maintenance using unmapped CLI tools, triggering novelty penalties.

\section{Limitations}

Although \textsc{AWS-IAM-ContextBench} provides controlled variation approximating enterprise-style IAM activity across multiple services, environments, and lineages, it does not reproduce the scale, organizational diversity, background entropy, or longitudinal drift of large production environments. All observations originate from a single controlled AWS testbed and do not establish transfer to independently sourced organizations or accounts. The supplementary structural analysis identifies 371 observations whose policy structure is absent from training folds, with $M_2$ achieving $\rho=0.9674$ and $\text{MAE}=1.2236$; however, this does not establish generalization to arbitrary policy structures outside the benchmark distribution. Similarly, the evaluation does not test zero-shot generalization to previously unseen behavioral patterns or attack sequences. CloudTrail delivery latency (5--15 minutes) also limits immediate real-time scoring. Additionally, the reference priority score is generated by an expert algorithmic oracle modeling analyst priorities rather than human panel consensus, measuring agreement with the defined scheme rather than human decision-making. Finally, evaluation is AWS-specific; while the conceptual formulation extends to cloud IAM generally, cross-provider validation remains future work.

\section{Conclusion}
This paper evaluated whether AWS CloudTrail behavioral context provides measurable value for interpretable IAM policy risk prioritization. Using \textsc{AWS-IAM-ContextBench}, we evaluated policy-centric, environment-augmented, and full-context Explainable Boosting Machines under a leakage-controlled grouped cross-validation protocol. The empirical findings indicate that incorporating CloudTrail behavioral telemetry reduces priority prediction error by 74.0\% ($p < 10^{-15}$) and increases rank correlation from $\rho = 0.4034$ to $\rho = 0.9726$. In matched same-policy pairs, the contextual model distinguishes suspicious from benign activity with 96.62\% direction accuracy ($p = 3.68 \times 10^{-30}$), while the policy baseline remains invariant. In simulated queues, context scoring yields a +19.7\% NDCG@10 lift. These results show that while authorization analysis establishes permission boundaries, behavioral telemetry provides essential evidence for analyst-oriented risk prioritization within an interpretable additive model architecture. Future work should evaluate the approach on independently sourced organizations and accounts, production-scale telemetry, structurally novel IAM policies, previously unseen behavioral patterns, continuous streaming models, human-subject SOC triage studies, and additional cloud providers.

\bibliographystyle{IEEEtran}
\bibliography{references}

\end{document}